\documentclass[nofootinbib,onecolumn,preprintnumbers,amsmath,amssymb,prd,showkeys]{revtex4}
\usepackage{amsmath,amsthm,amssymb}

\numberwithin{equation}{section}
\allowdisplaybreaks

\begin{document}
\title{Phenomenological covariant approach to gravity}
\preprint{IPM/P-2008/056}

\author{Qasem \surname{Exirifard}}
\affiliation{School of Physics, Institute for research in Fundamental Sciences (IPM), P.O.Box 19395-5531, Tehran, Iran}

\email{exir@theory.ipm.ac.ir}

\begin{abstract}
We covariantly modify the Einstein-Hilbert action such that the modified action perturbatively  resolves the anomalous rotational velocity curve of the spiral galaxies and gives rise to  the Tully-Fisher relation, and dynamically generates the cosmological constant. This modification requires introducing a single new universal parameter. It requires inclusion of neither dark matter nor dark energy.  
\end{abstract}
\maketitle
In this paper, we shall consider a Lagrangian density for the space-time geometry in the form of the Ricci scalar multiplied to an arbitrary functional of the Riemann tensor and its covariant derivatives. In the first section, we shall fix the form of the Lagrangian such that the modified Lagrangian resolves the missing mass problem in the galaxies without considering dark matter. This resolution is obtained by requiring that the spherical perturbative solution to the modified action for a point-like mass, coincides to the metric suggested by \cite{Sobouti:2009pa,Sobouti:2008et,Sobouti:2008xz}. This suggestion  takes into account the anomalous rotational velocity curves \cite{Rubin,Persic:1995ru,Borriello:2000rv,Salucci:2007tm} and the Tully-Fisher relation \cite{Tully}. We then shall prove that the modified Lagrangian dynamically generates the cosmological constant. All this procedure requires  introducing just a single new universal parameter. 

\section{The anomalous rotational velocity curve}
The space-time geometry around a stationary point-like mass distribution can be presented in the following coordinates:
\begin{equation}
ds^2= - A(r) dt^2 + B(r) dr^2 + r^2 d\Omega^2 \,.
\end{equation}
The dynamics of the space-time identifies $A(r)$ and $B(r)$ up to  the boundary conditions. In the Einstein-Hilbert gravity this metric in the asymptotically flat space-time geometry   coincides to the Schwarzschild geometry
\begin{equation}
A(r)\,=\, \frac{1}{B(r)}\,=\, 1 -\frac{r_h}{r} \,,
\end{equation}
where $r_h=\frac{2 G_N m}{c^2}$, $m$ is the mass, and $G_N$ is the Newton's gravitational constant. Ref. \cite{Sobouti:2009pa,Sobouti:2008et,Sobouti:2008xz} suggests that in order to account for the (almost) flat rotational velocity curve of the spiral galaxies, the space-time geometry around any point-like  distribution of observed matter deviates from the Schwarzschild geometry as follows:
\begin{equation}\label{Sobouti}
A(r)  \,=\, 1 -\frac{r_h}{r} + \epsilon \alpha r_h^{\frac{1}{2}} \ln r + O(\epsilon^2)\,,
\end{equation}
where $\epsilon$ is the systematic parameter of the perturbative series. Such a deviation can be  caused by the distribution of a kind of matter we could not directly observe, or it can be due to the deviation from the Einstein-Hilbert action. In this paper, we adhere to the later possibility. Such a possibility relies within the subject of modified gravities. Many studies have been conducted on $f(R)$-gravity \cite{Sotiriou:2008rp},  or $f(G)$-gravity \cite{Li:2007jm}. However neither of these classes of Lagrangians \cite{Exirifard:2007mb} admits the following perturbative solution:
\begin{subequations}\label{dark-matter-metrics}
 \begin{equation}\label{sch}
 ds^2 = - A(r) dt^2 + B(r) dr^2 + r^2 d\Omega^2 \,,
 \end{equation}
 \begin{equation}
A(r) = 1 -\frac{r_h}{r} + \epsilon a(r) + O(\epsilon^2)\,,
\end{equation}
\begin{equation}
B(r) = \,\frac{1}{A(r)} (1- 2 \epsilon b(r))\,
\end{equation}
\end{subequations}
where 
\begin{equation}\label{Suboutiar}
a(r)\,=\,\alpha r_h^{\frac{1}{2}} \ln(r)\,,
\end{equation} 
where $\alpha$ is a constant parameter.  We consider that the following action governs the dynamics of the space-time geometry
\begin{equation}
S \,=\, \frac{1}{8 \pi G_{N}} \int d^4x \sqrt{-\det g}\, F[R_{ijkl},\nabla,g_{\mu\nu}]\, R\,, 
\end{equation}
where $F[R_{ijkl},\nabla,g_{\mu\nu}]$ is a dimension-less scalar constructed out from the Riemann tensor and its covariant derivatives and $R$ is the Ricci scalar. This is an extension of $f(R)$ gravity. $F$ must have the following expansion
\begin{equation}\label{general_lagrangian}
 F[R_{ijkl,\nabla,g_{\mu\nu}}] = 1 + \epsilon L[R_{ijkl},\nabla,g_{\mu\nu}] + O(\epsilon^2)\,.
\end{equation} 
For the moment, we do not assume any constraint or form on $L[R_{ijkl},\nabla,g_{\mu\nu}]$. We consider it as an arbitrary functional of the Riemann tensor and its covariant derivatives. We aim to find the simplest form for $L[R_{ijkl},\nabla,g_{\mu\nu}]$ for which \eqref{dark-matter-metrics} is a perturbative solution. In the following, we  wish to first demonstrate how we will obtain the perturbative equations of motion.

Let the metric be presented as follows
\begin{equation}
g \,=\, g^{(0)} + \epsilon g^{(1)} + \cdots \,,
\end{equation}
where its indices are not shown but understood. We present the action as follows
\begin{equation}
S \,=\, \int d^4x \,(L^{(0)}[g] + \epsilon L^{(1)}[g] + \cdots) \,,
\end{equation}
where each $L^{(i)}$ is a functional of metric and its partial derivatives. Inserting the expansion of the metric into the expansion for the action we obtain:
\begin{eqnarray}\label{epsilon-action}
 S &=& \int d^4x L^{(0)}[g^{(0)}] + \epsilon \int d^4x (\left. \frac{\delta L^{(0)}}{\delta g} \right|_{\epsilon=0}.g^{(1)} + L^{(1)}[g^{(0)}])\,+ \nonumber\,\\
 && + \epsilon^2 \int d^4x \left(\left. \frac{\delta^2 L^{(0)}}{\delta^2 g} \right|_{\epsilon=0}.(g^{(1)})^2+ \left. \frac{\delta L^{(1)}}{\delta g} \right|_{\epsilon=0}. g^{(1)}\right)\,,
\end{eqnarray}
Note that the variation of the first term induces the equation of motion for $g^{(0)}$. These are the classical equations. We fix $g^{(0)}$ such that the classical equations are solved. Having done so the linear term in $\epsilon$ in \eqref{epsilon-action} becomes a constant number. We then take the variation of the quadratic terms in $\epsilon$ to obtain the equation of motion for $g^{(1)}$. This process gives a non-homogeneous linear equation for $g^{(1)}$. 

In the following, we apply the above procedure to the following action
\begin{equation}
S \,=\, \int d^4x\, \sqrt{-\det g} \,R\, (1 + \epsilon L[R_{ijkl}, \nabla_i,g_{ij}] )\,, 
\end{equation}
in order to find the equations of motion for perturbation around the Schwarzschild metric \eqref{dark-matter-metrics}. It is known that the functional variation does not generally commute with imposing symmetries on the solution; here we have imposed spherical symmetry and time translation. We notice, however, our imposed symmetries are isometries of the Riemann manifold -supposedly a smooth manifold- so the principle of symmetric criticality is met  \cite{Palais}. In other words the functional  variation of the  action  computed for \eqref{dark-matter-metrics},  indeed gives the equations of motion of $A(r)$ and $B(r)$. 

The non-vanishing components of the Ricci tensor calculated for \eqref{dark-matter-metrics} read\footnote{Note that  $R_{\mu\nu}=R^{\lambda}_{~\mu\lambda\nu}$ }:
 \begin{eqnarray}
R_{tt} &=& \frac{A''}{2 B} - \frac{A'}{4 B} (\frac{A'}{A}+\frac{B'}{B}) + \frac{A'}{r B}\,,\\
R_{rr} & =& -\frac{A''}{2 A} + \frac{A'}{4 A} (\frac{A'}{A}+\frac{B'}{B}) + \frac{B'}{r B}\,,\\
R_{\theta\theta} &=& 1 -\frac{1}{B} - \frac{r}{ 2 B} (\frac{A'}{A}-\frac{B'}{B})\,.
 \end{eqnarray}
Using the explicit forms of the Ricci tensor, it is straightforward to show that
\begin{equation}
\frac{1}{\sin \theta}\sqrt{-\det g} R = 0 + \epsilon \tilde{R}^{(1)} + \epsilon^2 \tilde{R}^{(2)} + \cdots \,,
\end{equation}
where
\begin{eqnarray}
\tilde{R}^{(1)}&=&-(r^2 a'' + 4 r b' - 3 r_h b' + 2 a + 4b + 4 r a')\,,\\
\tilde{R}^{(2)}&=& -4 r (a b)' - r^2 (a' b)' - 4 b^2 - 2ba + (9 r_h -12 r) b b'\,.
\end{eqnarray}
One also can  use algebraic tools like GrtensorII  to calculate the explicit form of $\tilde{R}^{(1)}$ and $\tilde{R}^{(2)}$ in term of $a(r)$ and $b(r)$. Since the Ricci scalar vanishes at the leading order, only the leading term in the $\epsilon$ expansion of $L[R_{ijkl}, \nabla_i,g_{ij}]$ contributes to the equations of motion of $a$ and $b$. We therefore replace $L[R_{ijk},\nabla_i,g_{ij}]$ with its value around the Schwarzschild metric
\begin{equation}\label{LSimple-LnotSimple}
L(r) =  L[R_{ijkl}, \nabla_i,g_{ij}]|_{\epsilon=0}\,.
\end{equation}
Doing so  the action reads
\begin{equation}\label{0-S}
S \,=\,4 \pi \int  dr dt  (0 + \epsilon \tilde{R}^{(1)} + \epsilon^2 
\tilde{R}^{(2)} +\cdots) (1 + \epsilon L(r) + O(\epsilon^2)) \,
\end{equation}
where integration on $\theta$ and $\phi$ are performed. The part of the action that contributes to the equations of motion of $a$ and $b$ then reads
\begin{equation}
S[a,b] \,=\, 4\pi \epsilon^2 \int dt \times \int dr \left(\tilde{R}^{(2)} + \tilde{R}^{(1)} L \right)\,, 
\end{equation}
Performing  integration by parts, the integration over $\tilde{R}^{(2)}$ is simplified to
\begin{equation}
\int dr \tilde{R}^{(2)} = \int dr (- 2 r b a' + 2 b^2) + cte\,,
\end{equation}
inserting which into \eqref{0-S} yields
\begin{equation}\label{2-S}
S[a,b] \,=\, 4\pi \epsilon^2\, \int dr dt \left(- 2 b' a r + 2 b^2 - L (r^2 a'' + 4b' r - 3r_h b' + 2 a + 4b + 4r a')\right)\,, 
\end{equation}
in which the explicit value of $\tilde{R}^{(1)}$ is also inserted . The first variation of \eqref{2-S} with respect to $a(r)$ and $b(r)$ generates the equations of motion for $a(r)$ and $b(r)$;
\begin{eqnarray}
 \frac{\delta S[a,b]}{\delta a} = 0 &\to& - 2 r b' - r^2 L''=0\,,\label{2-a}\\
 \frac{\delta S[a,b]}{\delta b} = 0 &\to& 2  a + 4 b + 2 r a' + (4 r-3 r_h) L'=0\label{2-b}\,.
\end{eqnarray}
We first solve \eqref{2-b} for $b(r)$
\begin{equation}
b(r) = c_1 + \frac{1}{2} (L - r L')\,.
\end{equation}
We then insert $b(r)$ into \eqref{2-a} and solve it for $L(r)$:
\begin{equation}\label{2-g-eq}
 L\,=\, -\frac{a(r)}{1 -\frac{3 r_h}{2r}} + \frac{4 r  c_1 + c_2}{3 r_h - 2r}\,,
 \end{equation}
 where $c_1$ and $c_2$ are constant of integration. In large $r$ ($r\gg r_h$), $L$ can be approximated to  
 \begin{equation}
 L\,\approx\, - a(r) -2 c_1- \frac{c_2}{2r}\,,
 \end{equation}
within which only $a(r)$ depends on the central mass distribution \eqref{Suboutiar} . $c_1$ and $c_2$ represent some effective cosmological parameters in the sense that these constants do not depend on the mass of the galaxy. Around a galaxy, presumably we can ignore the cosmological parameters. Therefore  we set $c_1=c_2=0$ and get
\begin{equation}
L(r) \approx \alpha r_h^{\frac{1}{2}} \ln r \,. 
\end{equation}
Noting  \eqref{LSimple-LnotSimple} then yields 
\begin{equation}\label{constraint_on_L}
L[R_{ijkl},\nabla_i,g_{ij}]|_{\epsilon=0} \approx \alpha r_h^{\frac{1}{2}} \ln r\,,
\end{equation}
which is an equation that helps us to find the form of $L[R_{ijkl},\nabla_i,g_{ij}]$. We would like to find the simplest form for $L[R_{ijkl},\nabla_i,g_{ij}]$ that satisfies \eqref{constraint_on_L}. In so doing we notice that
\begin{eqnarray}
 G|_{\epsilon=0}&=&\frac{12 r_h^2}{r^6}\,,\\
(\nabla G)^2|_{\epsilon=0}&=& \frac{5184 r_h^4 (r-r_h)}{r^{15}}\,\approx\,5184 \frac{r_h^4}{r^{14}}\,,\\
(\Box G)|_{\epsilon=0} &=& \frac{4 (90 r_h^2 r -108 r_h^3)}{r^9}\,\approx\, 360 \frac{r_h^2}{r^8}\,,
\end{eqnarray}
where $G$ stands for the Gauss-Bonnet term: 
\begin{equation}\label{GBterm}
G= R_{ijkl}R^{ijkl}-4R_{ij}R^{ij}+R^2\,.
\end{equation}
So we have
\begin{eqnarray}
\left(\frac{G}{(\Box G)^{\frac{3}{4}}}\right)_{\epsilon=0} &\approx& \frac{12}{(360)^{\frac{3}{4}}}\, r_h^{\frac{1}{2}}\,,\\
\ln (\frac{G}{\Box G}) |_{\epsilon=0}&\approx& 2 \ln r\,.
\end{eqnarray}
We therefore choose
\begin{equation}\label{choiceofc1}
 L[R_{ijkl},\nabla_i,g_{ij}]|_{\epsilon=0}\,=\, -\alpha_{_{DM}}\,\frac{G}{(\Box G)^{\frac{3}{4}}}\, \ln(c_1 \alpha_{_{DM}}^4\frac{ G}{\Box G})
\end{equation}
where $\alpha_{_{DM}}$ is a constant parameter and $c_1$ is a number \footnote{ Note that here we need $\Box G$ in the denominator to account for the Tully-Fisher relation. It is interesting that the existence of derivatives in the denominator of the Lagrangian is also required in the phenomenological approach to other problems in Cosmology  \cite{Woodard}.}. We set $c_1=1$. Also note that  $\alpha= \alpha_{_{DM}} \frac{24}{(360)^{3/4}}$. For this choice of $L[R_{ijkl},\nabla_i,g_{ij}]$ we find 
\begin{equation}\label{bfound}
b= -\frac{a}{2} -\frac{3 r_h^{\frac{1}{2}}\alpha}{r}\approx -\frac{a}{2}\,,
\end{equation}
Note that this value for $b$ means that the $rr$ component of the metric receives no leading correction, this is in agreement with the suggestion of \cite{Sobouti:2009pa,Sobouti:2008et,Sobouti:2008xz}. 
To summarize, we have found that 
\begin{equation}\label{aaa}
S\,=\, \frac{1}{8\pi G_N}\int d^4 x\, \sqrt{-\det g}\,R\, \left(1 -\alpha_{_{DM}} \frac{G}{(\Box G)^{\frac{3}{4}}}\,\ln(\alpha_{_{_{DM}}}^4\frac{ G}{\Box G})\right)  \,,
\end{equation}
by construction, admits the following perturbative solution
\begin{equation}\label{ultimate_metric}
ds^2 \,=\, -(1- \frac{r_h}{r} + \frac{24\alpha_{_{DM}}}{(360)^{\frac{3}{4}}} r_h^{\frac{1}{2}} \ln r+\cdots) dt^2 + (1+\frac{r_h}{r}+\cdots)dr^2+r^2 d\Omega^2\,. 
\end{equation}
Note that the $rr$ component of this metric is not the inverse of the $tt$ component of the metric.\footnote{The corrections to the equations of motion should be such that they do not have  vanishing radial null-null component \cite{Jacobson:2007tj}.} A classical (slow moving)  test probe in this geometry experiences the following effective gravitational acceleration (the leading gravitational potential is one half of the $tt$ component of the metric):
\begin{equation}\label{acceleration}
a_{\text{acc.}} =  \frac{c^2}{2} \frac{r_h}{r^2}  + \frac{12 \alpha_{_{DM}}c^2}{(360)^{\frac{3}{4}}} (\frac{r_h}{r^2})^{\frac{1}{2}}\,,
\end{equation}
the first term of which is the Newtonian gravitational acceleration. The second term becomes the dominant term when the Newtonian gravitational acceleration is smaller than 
\begin{equation}
a_{\text{critical}}\,=\, \frac{288  c^2}{(360)^{3/2}} \alpha_{_{DM}}^2 \,.
\end{equation}
In the MOND paradigm \cite{Milgrom:1983ca,Milgrom:1983pn}, the deviation from Newtonian gravitational dynamics is often expected beyond \cite{Bekenstein:2007iq,Begeman:1991iy}
\begin{equation}
a_0 = (1.2 \pm 0.27)\times 10^{-10}\frac{m}{s^2}\,,
\end{equation}
Inserting this value in the above relation results
\begin{equation}\label{valueofalpha}
\alpha_{_{DM}}\,=\, 1.78^{+0.19}_{-0.21} \times 10^{-13}\frac{1}{\text{meters}^\frac{1}{2}} \,.
\end{equation}
We would like to emphasize that \eqref{aaa} for the above value of $\alpha_{_{DM}}$ can resolve the anomalous rotational curves of the galaxies and also reproduce the Tully-Fisher relation \cite{Tully}.

We would like to address also the effective gravitational force of \eqref{ultimate_metric} to a relativistic (fast moving) massive object (or a probe). The exact effective potential for spacecrafts in the spherical and static geometry of \eqref{sch} is \cite{blau}
\begin{eqnarray}\label{Veff}
\frac{1}{2} \dot{r}^2 + V_{eff}(r) &=& 0\,,\\
V_{eff}(r)&=&\frac{l^2}{2 \,r^2\,B(r)}-\frac{E^2}{2\,c^2\, A(r)\, B(r)}+\frac{c^2}{2\,B(r)}\,,
\end{eqnarray}
where dot represents variation with respect to the proper time, and the equatorial plane is chosen to be orthogonal to the angular momentum, and $E$ stands for the energy (per unit rest mass) and $l$ represents the magnitude  (per unit rest mass) of the angular momentum. In an almost flat space-time geometry ($A \approx B \approx 1$) and for a slow moving object ($E\approx c^2$) and in large $r$, the effective potential can be approximated to 
\begin{eqnarray}\label{Veff1}
V_{eff}^{(0)}(r)&\approx&\frac{l^2}{2 \,r^2}+\frac{c^2(A(r)-1)}{2}\,,
\end{eqnarray}
which is square with saying that the gravitational potential at the leading order is one half of the $tt$ component of the metric. When $E$ is large ($E > c^2$) and the space-time geometry is almost flat,  the effective potential reads
\begin{eqnarray}\label{VeffE}
V_{eff}^{E}(r)&\approx&V_{eff}^{(0)}(r) - \frac{E^2 - c^4}{ 2 c^2 A(r) B(r)}\,,
\end{eqnarray}
Using the expansion series of $A$ and $B$ \eqref{dark-matter-metrics}, and using \eqref{bfound}, then leads to the following gravitational acceleration for a relativistic object
\begin{equation}\label{accelation_relativistic}
a_{\text{acc.}}^E \,=\, \frac{c^2}{2} \frac{r_h}{r^2}  + \epsilon \frac{E^2}{2 c^2} a'(r) + O(\epsilon^2, \epsilon \frac{m}{r})\,,
\end{equation}
For non-relativistic object $E\approx c^2$, \eqref{accelation_relativistic} converts to \eqref{acceleration}. The dependency on the velocity is due to the fact that $A(r) B(r) \neq 1$.   It implies that the gravitational force of the galaxy (or dark matter halo around the galaxy) effectively can be large for  massive objects with high speed $(v>\frac{c}{\sqrt{2}})$. So caution must be taken into account when the gravitational acceleration on objects with large speed is being addressed. The stars orbiting the galaxy have low speed $(\frac{v}{c}<<1)$. So \eqref{Veff} suffices to address the gravitational acceleration they experience.  

\section{Application to Dark Energy}
The cosmological constant in term of $h$ and $\Omega_\Lambda$ reads
\begin{equation}
\Lambda \, =\, \frac{3}{c^2}\, H_0^2 \Omega_\Lambda\,.
\end{equation}
Table IV of  \cite{Tegmark:2003ud} reports  values of $\Omega_\Lambda$  and $H_0$ due  to +SN Ia observation as :
\begin{eqnarray}
\Omega_\Lambda &=& 0.725^{+0.039}_{-0.044}\,,\\
h &=& 0.599^{+0.090}_{-0.062}\,,
\end{eqnarray} 
where $H_0 = 100\, h\, km\, s^{-1} MPc^{-1} = 3.24 \times 10 ^{-18} h s^{-1}$. Using these values, one finds
\begin{equation}\label{cosmologiacl_measured}
\Lambda_{\text{measured}} \,=\, 1.206^{+0.064}_{-0.073} \times 10^{-52} \frac{1}{\text{meters}^2} \,.
\end{equation} 
In the previous section we have shown that a universal parameter, $\alpha_{_{DM}}$, \eqref{valueofalpha} resolves the anomalous rotational velocity curves of the galaxies. We can utilize the value of  $\alpha_{_{DM}}$ to construct a unit for the cosmological constant:
\begin{equation}\label{alphaRC}
[\Lambda_{\alpha}] \,= \alpha_{_{DM}}^4 = 1.0^{+0.5}_{-0.4}  \times 10^{-51} \frac{1}{\text{meters}^2} .
\end{equation}
which is quite close to the measured value for the cosmological constant. In this section, we wish to modify the action for the dynamics of the space-time such that we account both for the anomalous rotational velocity curves of the galaxies and dark energy problem. 

Note that we add the cosmological constant to the Einstein-Hilbert action in order to have the following homogeneous solution\footnote{Note that the Cosmological constant is defined by $R_{ij}-\frac{1}{2} g_{ij} R= -\Lambda_{_\text{Measured}} g_{ij}$ which leads to \eqref{3-homo} where $\Lambda_{_\text{Measured}}=\Lambda$ is inferred. }:
\begin{eqnarray}\label{3-homo}
R^{a}_{~b}&=& \Lambda \delta^{a}_{~b}\,,\\
R &=& 4\,\Lambda\,,\nonumber\\
G &=& \frac{8}{3} \Lambda^2\,,\nonumber\\
\Box G &=&0\,,\nonumber
\end{eqnarray}
where $G$ stands for the Gauss-Bonnet term \eqref{GBterm}. The modification that we have presented in \eqref{aaa} diverges for the above solution. This divergence is due to the presence of $\Box G$ in the denominator of the modification. Let us replace $\Box G$ with $(\Box  + k R) G$ in \eqref{aaa} to obtain:
\begin{equation}\label{Action_Cosmological}
S\,=\, \int d^4 x\, \sqrt{-\det g}\,R\, \left(1 -\alpha_{_{DM}} \frac{G}{((\Box+k R) G)^{\frac{3}{4}}}\,\ln(\alpha_{_{{DM}}}^4\frac{ G}{(\Box+k R) G})\right)  \,,
\end{equation}
where $k$ is a non-vanishing number at order one while  $R$ is the Ricci scalar. Sufficiently close to a mass distribution where we can ignore the effects of the cosmological constant \eqref{Action_Cosmological} is the same as \eqref{aaa}. \eqref{Action_Cosmological}, however, admits non-trivial homogeneous solutions.

To find the exact non-trivial homogeneous solutions of \eqref{Action_Cosmological}, we must derive its equations of motion for metric.  In order to take the first variation of the action with respect to the metric (equations of motion), we first rewrite it as follows
\begin{eqnarray}
S &=&  \int d^4 x \sqrt{-\det g} \, {\cal L} \\
{\cal L} &=& {\cal L}[R, G , \Box G]\,=\, R\,\left(1 -\alpha_{_{DM}} \frac{G}{((\Box+k R) G)^{\frac{3}{4}}}\,\ln(\alpha_{_{{DM}}}^4\frac{ G}{(\Box+k R) G})\right)  \,.
\end{eqnarray}
The first variation of the action then reads
\begin{equation}\label{deltaS}
\delta S\,=\,  \int d^4x \,\sqrt{-\det g}\left(\frac{\partial {\cal L}}{\partial R} \delta R + (\frac{\partial {\cal L}}{\partial G}  + \Box \frac{\partial {\cal L}}{\partial \Box G})\delta G - \frac{1}{2} {\cal L} g_{ij} \delta g^{ij}\right)\,,
\end{equation}
The variation of the Ricci scalar reads
\begin{equation}\label{DR_value}
\delta R \,=\, R_{ij} \delta g^{ij} + 2 E^{R}_{ij\alpha\beta} \nabla^\alpha\nabla^\beta \delta g^{ij} \,,
\end{equation}
where $E^{R}_{ij\alpha\beta}$ is a tensor. The Gauss-Bonnet Lagrangian in four dimensions is topological. So in four dimensions, it holds
\begin{equation}\label{deltaG}
\int d^4 x \delta(G \sqrt{-\det g}) \,=\, 0 \,\to \, \delta G -\frac{1}{2} G  g_{ij} \delta g^{ij} + 2 E^{G}_{ij\alpha\beta}\nabla_{\alpha}\nabla_{\beta} \delta g^{ij}\,=\,0\,,
\end{equation}
where $E^{R}_{ij\alpha\beta}$ is a tensor. We shall not need the explicit forms of  $E^{R}_{ij\alpha\beta}$ and $E^{R}_{ij\alpha\beta}$. We, therefore, don't write them. \eqref{deltaG} leads to 
\begin{equation}\label{ValueOfDG}
\delta G \,=\, \frac{1}{2} G  g_{ij} \delta g^{ij} - 2 E^{G}_{ij\alpha\beta}\nabla_{\alpha}\nabla_{\beta} \delta g^{ij}\,.
\end{equation}
Inserting \eqref{ValueOfDG} and \eqref{DR_value} in \eqref{deltaS} leads to the equations of motion
\begin{equation}\label{Eq00}
\frac{\partial {\cal L}}{\partial R}  R_{ij} + \frac{G}{2}(\frac{\partial {\cal L}}{\partial G}  + \Box \frac{\partial {\cal L}}{\partial \Box G})g_{ij} - \frac{1}{2} {\cal L} g_{ij}  + 2 \nabla^{\alpha}\nabla^{\beta} (\frac{\partial {\cal L}}{\partial R} E^{R}_{\alpha\beta i j} - (\frac{\partial {\cal L}}{\partial G}+ \Box \frac{\partial {\cal L}}{\partial\Box  G} ) E^G_{\alpha\beta i j}) \,=\,0\,.
\end{equation}
We would like to find the following solution of the above equations (a dS space-time geometry):
\begin{equation}\label{dS}
ds^2 \,=\, - (1- \frac{\Lambda}{3} r^2 ) dt^2+ \frac{dr^2}{1- \frac{\Lambda}{3} r^2 } + r^2 (d\theta^2 + \sin^2 \theta d\phi^2)
\end{equation}
For this solution, the Riemann tensor is covariantly constant:
\begin{equation}
\nabla_a R_{ijkl}\,=\,0\,.
\end{equation} 
Therefore any tensor constructed out from the Riemann tensor and its covariant derivatives are covariantly constant. Due to this reason, the equations of motion \eqref{Eq00} calculated for  \eqref{dS} are simplified to 
\begin{equation}\label{EqdS0}
\frac{\partial {\cal L}}{\partial R}  R_{ij} + \frac{G}{2}\frac{\partial {\cal L}}{\partial G} g_{ij} - \frac{1}{2} {\cal L} g_{ij}   \,=\,0\,.
\end{equation}
Furthermore we note that \eqref{dS} holds 
\begin{equation}
R_{ij}\,=\, \Lambda g_{ij}
\end{equation}
inserting  which in \eqref{EqdS0} yields
\begin{equation}\label{EqdS01}
(\frac{\partial {\cal L}}{\partial R} \Lambda + \frac{G}{2}\frac{\partial {\cal L}}{\partial G}  - \frac{1}{2} {\cal L})g_{ij} \,=\,0\,,
\end{equation}
which can be simplified to
\begin{equation}\label{EqdS1}
\frac{\partial {\cal L}}{\partial R} \Lambda + \frac{G}{2}\frac{\partial {\cal L}}{\partial G}  - \frac{1}{2} {\cal L}  \,=\,0\,.
\end{equation} 
When the value of $\Lambda$ is such that  \eqref{EqdS1} holds than \eqref{dS} is an exact solution of  \eqref{Action_Cosmological}. After taking the variation of the Lagrangian density with respect to $R$ and $G$ and using $R= 4 \Lambda$ and $G= \frac{8}{3}\Lambda^2$, \eqref{EqdS1} leads to the equation of motion for $\Lambda$. In order to succinctly express this equation we express $\Lambda$ in term of a new variable defined below
\begin{equation}\label{Lambda-K}
\Lambda \,\equiv\, \frac{\alpha_{_{DM}}^4}{4k} x^4\,. 
\end{equation}
The equation for $\Lambda$ then gives
\begin{equation}\label{xeq00}
 \sqrt{|k|} x + 6^{-\frac{1}{4}}(-1+ 5 \ln x)=0\,,
\end{equation}
This equation has a (real) solution provided that $-1+5 \ln x < 0$. So we set the domain of the $x$ variable to $x\in[0,e^{\frac{1}{5}}]$. In this domain, \eqref{xeq00} can be solved for $k$;
\begin{equation}\label{kx}
k\,=\, \frac{(1-5 \ln x)^2}{\sqrt{6}x^2}\,. 
\end{equation}
Also note that we have chosen $k$ to be positive, because $x\in[0,e^{\frac{1}{5}}]$ and we want a positive $\Lambda$.  Inserting \eqref{kx} in \eqref{Lambda-K} expresses $\Lambda$ solely in term of the $x$ variable:
\begin{equation}
\Lambda \,=\,\alpha_{_{DM}}^4 \frac{\sqrt{6}\, x^6}{4(1-5 \ln x)^2} \,. 
\end{equation}
We require $\Lambda=\Lambda_{_\text{Measured}}$ while $\alpha_{DM}$ approximately reads in \eqref{valueofalpha}. So
\begin{eqnarray}
\frac{\Lambda}{\alpha_{_{DM}}^4} &=& \frac{1.2 \times 10^{-52}}{(1.78 \times 10^{-13})^4}~ \to
\\ \label{eqx}
  0.1195 &=& \frac{\sqrt{6}\, x^6}{4(1-5 \ln x)^2} \,. 
\end{eqnarray}
Note that \eqref{eqx} is solved by $x=0.8892$. We, however, want to identify $k$. The only condition on $k$ is that it should be at order one. But we  require $k$ to be also either a natural or the inverse of a natural number. According to \eqref{kx}, $x=0.9167$ leads to $k=1$. This value of $x$ is not far from the value that solves \eqref{eqx}.\footnote{$x=0.8415$ leads to $k=2$. $x=0.8892$ yields $k=1.3$. So $k=1$ is the best choice.}  So we set $k=1$ and $x=0.9167$. For these values, the effective cosmological constant reads 
\begin{equation}\label{lambdaAlpha}
\Lambda \,=\, 0.1766\, \alpha_{_{DM}}^4 \,.
\end{equation}
Now we can read the best value of $\alpha_{_{DM}}$ from \eqref{lambdaAlpha} and $\Lambda=\Lambda_{_\text{Measured}}$:
\begin{equation}\label{alpha_finale}
\alpha_{_{DM}}\,=\, 1.617^{{+0.021}}_{{-0.025}} \times 10^{-13} \frac{1}{\text{meters}^\frac{1}{2}}\,,
\end{equation}
which is in agreement with \eqref{alphaRC} identified by the critical acceleration in the MOND paradigms. So in summary
\begin{equation}\label{Action_Cosmological_finale}
S\,=\, \int d^4 x\, \sqrt{-\det g}\,R\, \left(1 -\alpha_{_{DM}} \frac{G}{((\Box+ R) G)^{\frac{3}{4}}}\,\ln(\alpha_{_{DM}}^4\frac{ G}{(\Box+  R) G})\right)  \,,
\end{equation}
 includes only a single parameter, and   dynamically generates the measured value of the cosmological constant, perturbatively resolves the anomalous rotational velocity curves of the galaxies and gives rise to the the Tully-Fisher relation. 

\section{Summary and outlook}
We have shown that a covariant modification to the Einstein-Hilbert action can resolve the missing mass problem in galaxies, gives rise to the Tully-Fisher relation, and exactly the same modification dynamically generates the cosmological constant. We have constructed the simplest form for the action that is capable of performing these tasks \eqref{Action_Cosmological_finale}. This modification has one single new universal parameter, $\alpha_{_{DM}}$ \eqref{alpha_finale}. So the suggested modified action unifies the missing mass problem in the galaxies and dark energy problem before resolving both of them. 

Our model is not the first geometric model to unify and resolve dark matter and dark energy problems \cite{Capozziello:2006uv,Capozziello:2006ph, Boehmer:2007kx}. The previous resolutions, however, do not take into account the Tully-Fisher relation. The metric that they assign to a galaxy has a free parameter and that free parameter depends on  the baryonic mass of the galaxy. In comparison with these, our model dynamically generates the Tully-Fisher relation, and describes the space-time geometry around a galaxy with an arbitrary mass only with a universal parameter.  Its action, however, seems more complicated than the previous models. However, since it needs only one parameter to describe various phenomena we tend to evaluate it `physically simple'. 

The covariant resolution we suggest here, however, should be still counted on as a toy model because the ultimate covariant resolution of the dark energy and dark matter problems must address  all the phenomena assigned to dark matter such as the weak gravitational lensing, the bullet cluster \cite{BulletCluster}, dynamics of clusters of galaxies, and the CMB data. The stability of the cosmological perturbations also must be addressed. Let it be emphasized that, however, each of these phenomena demands a set of finite conditions on the form of the Lagrangian governing the dynamics of the space-time. Since there exist infinite possibilities available for the form of the Lagrangian, one expects to find  a set of Lagrangians that meets all these conditions.  In this paper we already have illustrated that seemingly separated phenomena such as cosmological constant and the missing mass problem in galaxies can be due to a single cause. We hope that further investigations render that some additional phenomena assigned to dark matter and dark energy can also be stemmed from a common single cause: a covariant modified Lagrangian for the dynamics of the space-time.

\providecommand{\href}[2]{#2}\begingroup\raggedright

\end{document}